\documentclass[twocolumn]{svjour3}          
\smartqed  
\usepackage{graphicx}
\usepackage{amsmath,amssymb}
\begin{document}

\title{Solitons and instantons in vacuum stability: physical phenomena
}


\author{J. A. Gonz\'{a}lez         \and A. Bellor\'{\i}n \and L. E. Guerrero \and S. Jim\'enez \and Juan F. Mar\'in \and L. V\'azquez
}


\institute{J. A. Gonz\'{a}lez \at
              Department of Physics, Florida International University, Miami, Florida 33199, USA \\
\email{jorgalbert3047@hotmail.com}           
           \and
           A. Bellor\'{\i}n \at
            Escuela de F\'{\i}sica, Facultad de Ciencias, Universidad Central de Venezuela, Apartado Postal 47586, Caracas 1041-A, Venezuela \\
\email{alberto.bellorin@ucv.ve}  
           \and
           L. E. Guerrero \at
           Departamento de F\'{\i}sica, Universidad Sim\'{o}n Bol\'{\i}var, Apartado 89000, Caracas 1080-A, Venezuela \\
           \email{lguerre@usb.ve}
           \and
           S. Jim\'enez \at
           Departamento de Matem\'{a}tica Aplicada a las TIC, ETSI Telecomunicaci\'{o}n, Universidad Polit\'{e}cnica de Madrid, 28040-Madrid, Spain \\
           \email{s.jimenez@upm.es}
           \and
           J. F. Mar\'in \at Departamento de F\'isica, Universidad de Santiago de Chile, Usach, Av. Ecuador 3493, Estaci\'on Central, Santiago, Chile.\\
          \email{juan.marin.m@usach.cl}
           \and
           L. V\'azquez \at
           Departamento de Matem\'atica Aplicada, Facultad de Inform\'atica, Universidad Complutense de Madrid, 28040-Madrid, Spain\\
          \email{lvazquez@fdi.ucm.es}
}

\date{Received: 27 May 2020 / Accepted: 26 August 2020}

\maketitle

\begin{abstract}
\noindent In a previous paper [JCAP06, 033 (2018)] we have proved that it is possible to have a stable false vacuum in a potential that is unbounded from below. In this paper we discuss the Physics related to our theoretical and numerical results. We show that the results of recent CERN experiments lead to the fact that our vacuum is safe. We present a new mechanism, where the space-time dimension plays an important role, that explains why our Universe is stable. We provide new evidence that supports a process for the origin of matter-antimatter asymmetry recently introduced by other scientists. We examine confinement in the context of escape problems. We discuss multiverse, string theory landscape, and extra-dimensions using our framework. Finally, we use our solutions to introduce some hypotheses about Dark Matter and Dark Energy.
\end{abstract}

%
%

\section{Introduction}
\label{intro}

First-order phase transition is one of the most important problems in condensed  matter, particle physics, and cosmology \cite{Cohen1993,Trodden1999,Yagi2005,Linde1990,Weinberg2008,Hanggi1988,Marin2018,GarciaNustes2017,Aubry1975,Marchesoni1998}. This field has applications in vacuum stability \cite{Cohen1993}, quark confinement \cite{Yagi2005}, and cosmological models \cite{Kibble1980}. The physics of this phenomenon has long been discussed in the literature \cite{Shkerin2015,Branchina2015,Khan2014,Ge2016,Kusenko2015,Hook2015,Bednyakov2015,Langer1967,Kusenko2015b,Coleman1977,Callan1977,Coleman1980,Kobzarev1975,Gorsky2015,Blum2015,Isidori2001,Turner1982,Lindner1989,Sher1989,Krive1976,Cabibbo1979}. It  is very similar to the nucleation processes of statistical physics, the crystallization of supersaturated solution or the boiling of superheated fluid.

\begin{figure}[tbh]
\centerline{\includegraphics[width=3.5in]{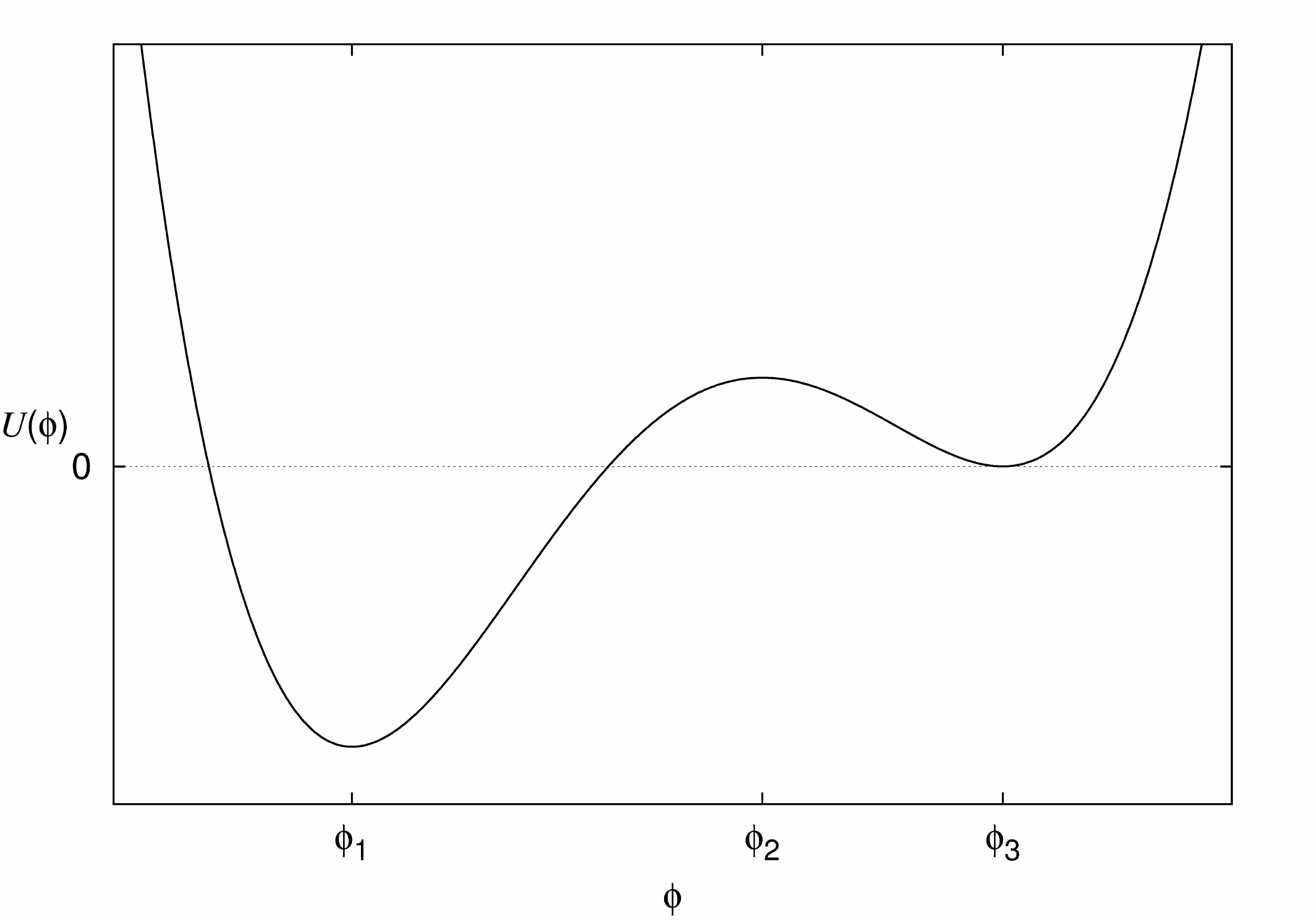}}
\caption{Potential $U(\phi)$ for a first-order phase transition.}
\label{fig:1}
\end{figure}

Suppose Fig.~\ref{fig:1} represents the free energy of a fluid as a function of density. The phase $\phi = \phi_3$ corresponds to the superheated fluid phase and $\phi = \phi_1$ to the vapor phase. Thermodynamic processes are continually causing bubbles of the vapor phase to materialize in the fluid phase. If the bubble is too small, it will shrink and will disappear. On the other hand, if the formed bubble is large enough, there are thermodynamic conditions energetically favorable for the bubble to grow. This bubble will expand until it converts the available fluid to vapor \cite{Langer1967}.

The generation and stability of bubbles are also relevant in cosmological models of the fate of the Universe \cite{Kusenko2015b,Coleman1977,Callan1977,Coleman1980,Kobzarev1975,Souza2020}. According to the Standard Model (SM), the recently measured masses of the Higgs boson's and top quark suggest that our Universe is currently in a metastable state of the Higgs potential \cite{Degrassi2012,Burda2016,EliasMiro2012}. Such kind of metastable states are termed as \emph{false vacuum states} and are similar to the phase $\phi=\phi_3$ in the potential of Fig.~\ref{fig:1}. Therefore, due to quantum fluctuations, a phase transition may occur from the current false vacuum to a \emph{true vacuum state} -- a global minimum in the Higgs potential. In such a new vacuum state, which is similar to the phase $\phi=\phi_1$ in Fig.~\ref{fig:1}, the way in which elementary particles interact would be completely different. A transition of the universe towards a nearby global minimum of the Higgs potential is usually termed as \emph{vacuum decay} and would lead the Universe towards a state where life as we know it might be impossible \cite{Turner1982}.

In this article, we discuss a possible scenario where a stable false vacuum state can be stable even if there is a true vacuum state nearby. True-vacuum bubbles created by quantum fluctuations would not grow, and there will be no bubble expansion that would convert false vacuum into true vacuum. Our findings lead us to a series of conjectures on the actual vacuum lifetime and suggest that such a phase transition would be highly unlikely. Thus, our current vacuum would be safe.

The remainder of the article is organized as follows. In Sections \ref{model} and \ref{instanton}, we recall our previous mathematical results \cite{Gonzalez2018}: in Section \ref{model}, we describe the system in the framework of vacuum decay in the Higgs potential, whereas in Section \ref{instanton}, we calculate the instanton solution connecting the vacua of a proposed potential, and we show that the action of the instanton can be divergent if some conditions are provided. In Section \ref{Numerical} we explore numerically in detail the collapse of a bubble of true vacuum inside a sea of false vacuum and show
that there are no bubbles of true vacuum inside a Universe of false vacuum that will expand. In Section \ref{Discussion} we discuss related physical phenomena and we introduce some hypotheses about current open questions. Finally, we conclude in Section \ref{Conclusions}.

\section{The system: Vacuum decay in the Higgs potential}
\label{model}

\begin{figure*}[h]
\centerline{\includegraphics[width=2.7in]{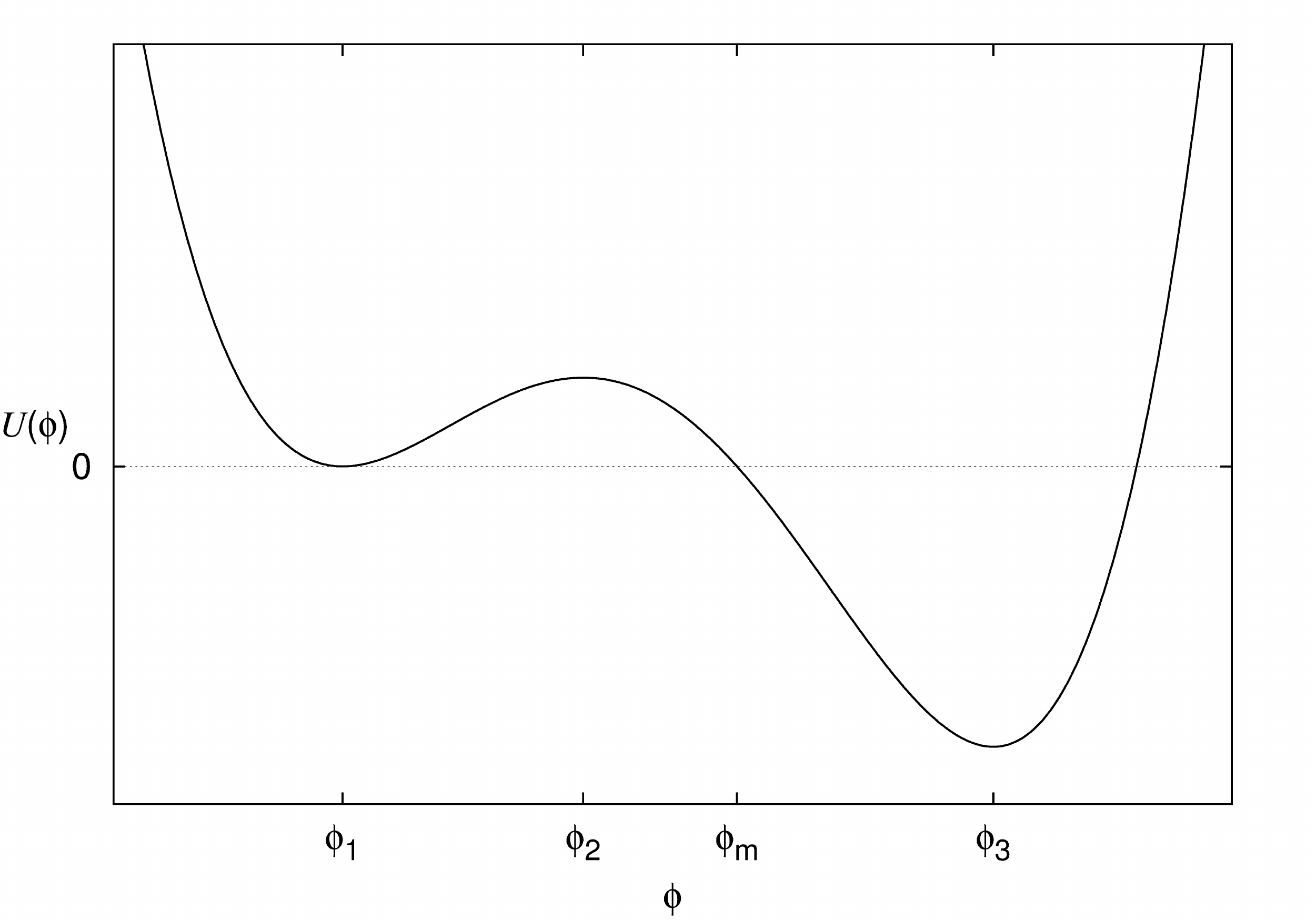}
\includegraphics[width=2.7in]{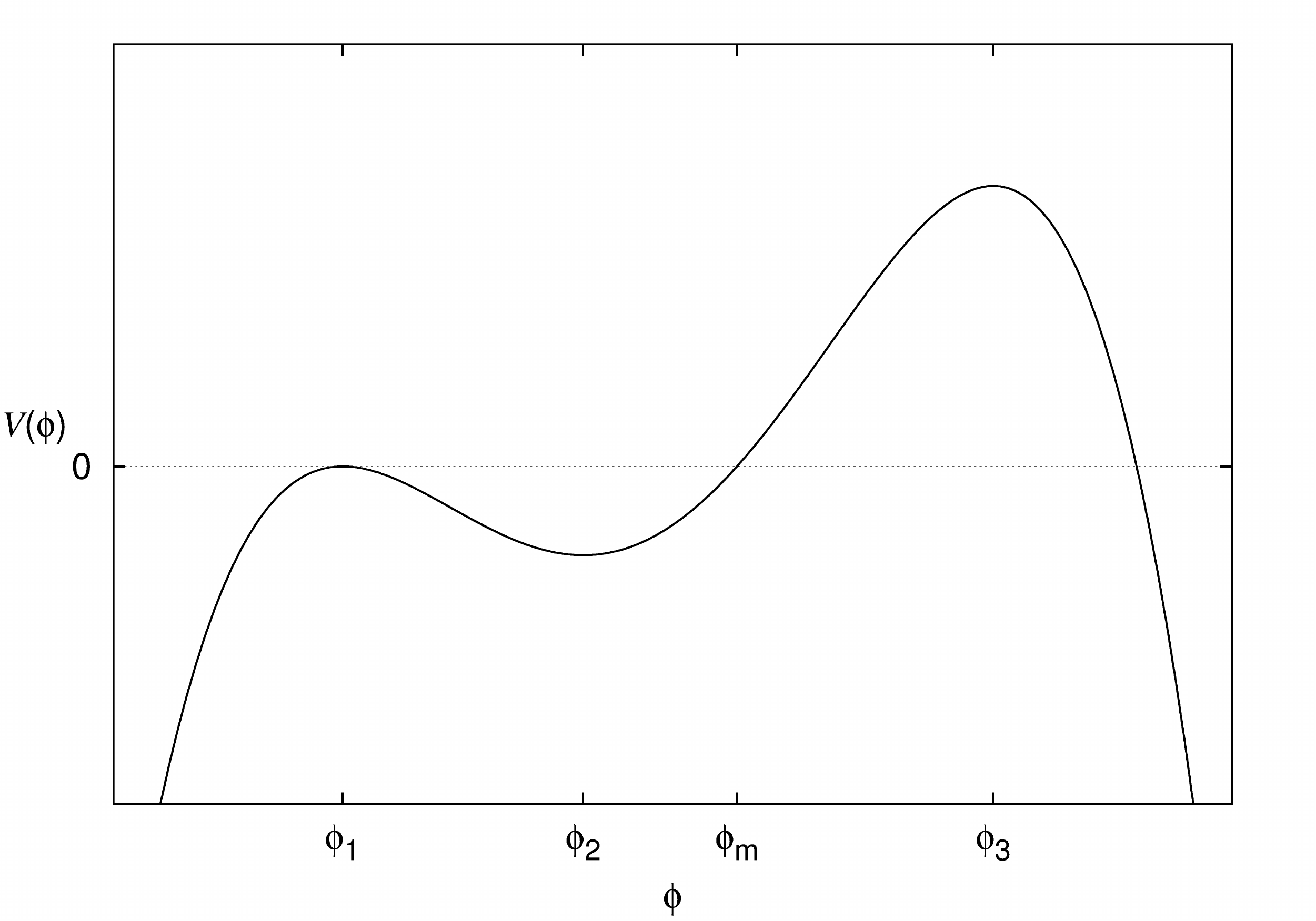}}
\caption{\textbf{(a)} Potential $U(\phi)$ with a false vacuum state $\phi_1$ and a true vacuum state $\phi_3$. Vacuum decay corresponds to instantons solutions to Eq.~\ref{Eq04} interpolating $\phi_1$ and $\phi_3$. \textbf{(b)} A fictitious particle moving in the potential $V(\phi)=-U(\phi)$. For the instanton to exist, the particle should pass the bottom of the valley $V(\phi_2)$ and come to rest at time infinity on top of the maximum $V(\phi_1)=0$.}
\label{fig:2}
\end{figure*}

Let us begin considering a general Lagrange density
\begin{equation}
\mathcal{L}=\frac{1}{2}\left(\partial_{\mu}\phi\right)\left(\partial^{\mu}\phi\right)-U(\phi),
\end{equation}
where $\phi$ is a real scalar field in $D$ dimensions and $U(\phi)$ is a potential whose graphical representation is shown in Fig.~\ref{fig:2}(a). It possesses two non-degenerate minima $\phi_1$, $\phi_3$, and a maximum $\phi_2$ such that $\phi_1<\phi_2<\phi_3$ and $U(\phi_3)<U(\phi_1)$. The lower of the minima, at $\phi=\phi_3$, is the true vacuum state, whereas the higher minimum, the state $\phi=\phi_1$, is a false vacuum state. We restrict ourselves to the case of zero temperature in flat space-time, and we define the zero-energy level at the false vacuum state, i.e. $U(\phi_1)=0$.

The false vacuum is unstable by quantum effects, such as barrier penetration. Quantum fluctuations are continually creating small regions of true vacuum in the Universe-- the so-called \emph{true-vacuum bubbles} --which are immersed in a false vacuum background. A bubble of true vacuum can eventually be formed large enough so that it is energetically favorable for the bubble to grow. For the vacuum to decay in the potential of Fig.~\ref{fig:2}(a), it must tunnel through the energy barrier at $\phi=\phi_2$. The probability of vacuum decay per unit time and per unit volume has an exponential factor \cite{Coleman1977,Callan1977,Coleman1980,Kobzarev1975}, $\Gamma/V\propto e^{-\beta/\hslash}$, where $\beta$ is the Euclidean action
\begin{equation}
 \label{Eq07}
 \beta=\int\,d\tau d^3x\left[\frac{1}{2}\left(\partial_{\tau}\phi\right)^2+\frac{1}{2}\left(\boldsymbol{\nabla}\phi\right)^2+U(\phi)\right],
\end{equation}
and $\tau:=it$ is the imaginary time. The corresponding classical Euclidean equation of motion is
 \begin{equation}
  \label{Eq04}
  \partial_{\tau\tau}\phi+\nabla^2\phi=\frac{dU}{d\phi}.
 \end{equation}
Solutions of Eq.~(\ref{Eq04}) have a finite action $\beta$ only if \cite{Weinberg2012}
$$\lim_{|\mathbf{x}|\to\infty}\phi=\phi_1.$$

If vacuum decay happens, a bubble of true vacuum expands throughout the Universe converting false vacuum to true. Such decay corresponds to instanton solutions of Eq.~(\ref{Eq04}) which interpolates between the false a true vacua with finite and non-vanishing action \cite{Weinberg2012}. Imposing the finiteness of the Euclidean action (\ref{Eq07}), the following conditions must be fulfilled \cite{Weinberg2012}
\begin{eqnarray}
\label{Eq05}
  \partial_{\tau}\phi(\tau=0, \mathbf{x})& = & 0,\\
\label{Eq06}
  \lim_{\tau\to\pm\infty}\phi(\tau,\mathbf{x}) & = &   \phi_1.
\end{eqnarray}

Given that the instanton solution is invariant under four-dimensional Euclidean rotations, it is possible to define $\rho:=\sqrt{|\mathbf{x}|^2+\tau^2}$ and seek for an instanton solution that depends only on $\rho$. From Eqs.~\eqref{Eq07}--\eqref{Eq04} we obtain
\begin{equation}
 \label{Eq13}
 \beta=2\pi^2\int_0^{\infty}d\rho\,\rho^3\left[\frac{1}{2}\left(\frac{d\phi}{d\rho}\right)^2+U(\phi)\right],
\end{equation}
\begin{equation}
\label{Eq10}
 \frac{d^2\phi}{d\rho^2}+\frac{3}{\rho}\frac{d\phi}{d\rho}=-\frac{dV(\phi)}{d\phi},
\end{equation}
where $V(\phi)=-U(\phi)$. Following Eqs.~(\ref{Eq05})--(\ref{Eq06}), the instanton solutions exist if
\begin{eqnarray}
\label{Eq11}
  \frac{d\phi(0)}{d\rho}& = & 0,\\
\label{Eq12}
  \lim_{\rho\to\infty}\phi(\rho) & = &   \phi_1.
\end{eqnarray}

The existence of a solution to Eqs.~\eqref{Eq10}-\eqref{Eq12} is commonly demonstrated by Coleman's well-known \emph{overshoot}-\emph{undershoot} argument \cite{Coleman1977}.  Following Coleman's approach, Eq.~\eqref{Eq10} is equivalent to Newton's equation of motion in one dimension for a fictitious particle of unit mass with ``pseudo-position" $\phi$ and ``pseudo-time" $\rho$. The particle is under the influence of the potential $V(\phi)$ shown in Fig.~\ref{fig:2}(b) and a dissipative force that decreases in pseudo-time
\begin{equation}
 \label{Eq14}
 F_{dis}=-\frac{3}{\rho}\frac{d\phi}{d\rho}.
\end{equation}

In false vacuum decay, we are interested in solutions where the fictitious particle starts at rest at some initial position $\phi(0):=\phi_{0}$ on the true-vacuum side of the potential $U(\phi)$, i.e.  $\phi_m<\phi_0<\phi_3$. For the instanton solution to exist, the particle should pass the minimum point $\phi=\phi_2$ and come to rest at time infinity on top of the maximum $V(\phi_1)$, as depicted in Fig.~\ref{fig:2}(b). If the potential $V(\phi)$ behaves as $V(\phi)\sim(\phi_2-\phi)^2/2$ in a neighborhood of point $\phi=\phi_2$, we can expect that the particle can make oscillations around such a fixed point. In the one-dimensional case, the dissipative force of Eq.~\eqref{Eq14} vanishes and oscillations around $\phi_2$ are sustained. For the $D=3+1$ case, the particle can make damped oscillations around the point $\phi_2$. Moreover, such oscillations can be in the over-damped regime for some potentials, and a particle starting its motion at a point $\phi_0>\phi_2$ will follow an over-damped trajectory that will end at point $\phi_2$ for $\rho\to\infty$, never visiting the region $\phi<\phi_2$. Nevertheless, in all these cases the instanton solution can still exist.

\section{The instanton solution}
\label{instanton}

In many articles, the instanton solutions are usually obtained using the so-called thin-wall approximation \cite{Coleman1977} and the $\phi^4$ theory as a model. The employed solution is nothing else but the one-dimensional $\phi^4$ soliton. However, such an approach loses some complexities of the 3+1 dimensional nonlinear field equations. In this section we will show that for some potentials, the action of the instanton solution can be arbitrarily large.

Let us briefly consider a potential with a single barrier at $\phi_2$ separating a vacuum $\phi_1$ and an abyss, i.e. a potential that is unbounded from below. Suppose that such potential has a quadratic (quartic) behavior around its minimum (maximum). Such a potential can be written as
\begin{equation}
 \label{Eq15}
 U(\phi)=\left\{\begin{array}{lcc}
 \displaystyle\frac{1}{2}a_1(\phi-\phi_1)^2, & \text{for} & \phi\leq\phi_{12},\\
 \displaystyle-\frac{1}{4}(\phi-\phi_2)^4+\Delta_1, & \text{for} & \phi\geq\phi_{12},
 \end{array}\right.
\end{equation}
where $a_1$ is a constant and $\Delta_1$ is the height of the barrier at $\phi=\phi_2$. We have introduced in Eq.~\eqref{Eq15} the intermediate point $\phi=\phi_{12}$ where both parts of the potential are smooth and continuously joint \footnote{The intermediate point $\phi_{12}$ can be obtained solving $a_1(\phi_{12}-\phi_1)^2/2 = \Delta_2-(\phi_{12}-\phi_2)^4/4$, and
$a_1(\phi_{12}-\phi_1)=-(\phi_{12}-\phi_2)^3$.}. All the solutions of Eq.~\eqref{Eq10} with the conditions
\begin{equation}
\label{Eq16}
\phi(0)>\phi_2,\quad
\left.\frac{d\phi}{d\rho}\right|_{\rho=0}=0,
\end{equation}
are given by the family of functions
\begin{equation}
 \label{Eq17}
 \phi(\rho)=\phi_2+Q\left(1+\frac{Q^2}{8}\rho^2\right)^{-1},
\end{equation}
where $Q>0$ is the distance of the initial condition from the vacuum ($\phi(0)=\phi_2+Q$). This includes solutions with the initial conditions
\begin{equation}
\label{Eq18}
 \phi(0)>\phi_{\text{m}},\quad 
 \left.\frac{d\phi}{d\rho}\right|_{\rho=0}=0.
\end{equation}

Note that there exists a continuum of solutions provided by Eq.~\eqref{Eq17}. For every initial condition $\phi(0)>\phi_2$, there is a solution in the family \eqref{Eq17}. Thus, returning to the equivalent system of a fictitious particle in the potential $V(\phi)=-U(\phi)$, we see that the particle will be trapped at $\phi=\phi_2$ even for $\rho\to\infty$. It does not matter how high the fictitious particle is at $\rho=0$, it will never visit points close to $\phi=\phi_1$. From Eq.~\eqref{Eq12} follows that the action of all these solutions is infinite.

In general, for any potential that behaves as $U(\phi)\simeq-(\phi-\phi_2)^4/4$ in a neighborhood of $\phi=\phi_2$, the fictitious particle for $D=3+1$ will be difficult near the barrier at $\phi_2$. For example, if we now consider a potential with a true-vacuum state after the barrier,
\begin{equation}
\label{Eq19}
  U(\phi)=\left\{\begin{array}{lcc}
 \displaystyle\frac{1}{2}a_1(\phi-\phi_1)^2, & \text{for} & \phi<\phi_{12},\\
 \displaystyle-\frac{1}{4}(\phi-\phi_2)^4+\Delta_1, & \text{for} & \phi_{12}<\phi<\phi_3,\\
 \displaystyle a_2(\phi-\phi_3)-\Delta_2, & \text{for} & \phi>\phi_3,
 \end{array}\right.
\end{equation}
where $a_2$ is a constant and $\Delta_2$ is the depth of the true vacuum state at $\phi_3$, there are no initial positions
$\phi_2<\phi(0)<\phi_3$ with $d\phi(0)/d\rho=0$ for which Eq.~\eqref{Eq12} is fulfilled.

Now let us consider the generalization of Eq.~\eqref{Eq10} in a $D$-dimensional space-time
\begin{equation}
 \label{Eq20}
 \frac{d^2\phi}{d\rho^2}+\frac{D-1}{\rho}\frac{d\phi}{d\rho}=-\,\frac{d V(\phi)}{d\phi},
\end{equation}
with the following potential
\begin{equation}
\label{Eq21}
  U(\phi)=-V(\phi)=\frac{a|\phi|^3}{3}-\frac{b|\phi|^4}{4},\quad\quad a>0,\,b>0.
\end{equation}
This potential is unbounded from below and has a well at $\phi_1=0$ between two symmetric barriers at $\pm\phi_2=\pm a/b$. The exact instanton solution to \eqref{Eq20}-\eqref{Eq21} is
\begin{subequations}
\label{Eq23}
\begin{align}
\label{Eq22}
 \phi_{\text{inst}}(\rho)&=\frac{Q}{1+N\rho^2},\\
\label{Eq23b}
\text{where}\quad Q:=\frac{4a}{(4-D)b}\,&,\quad
 N:=\frac{2a^2}{(4-D)^2b}.
 \end{align}
\end{subequations}

Notice that the instanton solution (\ref{Eq22}) is unique in the sense that there is only one initial condition $\phi(0)=Q$ for which
$\lim_{\rho\to\infty}\phi(\rho)=\phi_1=0$. In the framework of the fictitious particle moving in the potential $V(\phi)=-U(\phi)$, the instanton solution corresponds to the damped trajectory of a particle released a distance $Q$ from the barrier, reaching the top of the potential at rest and asymptotically in time. However, as $D$ is increased from $D=1$, the dissipative force $F_{dis}=(D-1)\rho^{-1}d\phi/d\rho$ will increase and the particle must be released from a larger distance $Q$ to have enough initial potential energy to be able to climb the hill and reach the point $\phi=\phi_1$. Thus, as $D$ is increased from $D=1$, the ``amplitude'' $Q$ of the instanton solution will increase.

However, it is important to remark that $Q\to\infty$ as $D\to4$. Thus, for $D=4$, it is very difficult for the fictitious particle to arrive at point $\phi=\phi_1=0$. This can be also an evidence that the action $\beta_{\text{inst}}$ of the instanton solution is such that
 \begin{equation}
  \label{Eq24}
  \lim_{D\to4}\beta_{\text{inst}}=\infty.
 \end{equation}
For instance, if we now consider the potential
\begin{equation}
\label{Eq25}
  U(\phi)=\left\{\begin{array}{lcc}
 \displaystyle\frac{1}{3}|\phi|^3-\frac{1}{4}|\phi|^4, & \text{for} & \phi<\phi_3,\\
 \displaystyle a_2(\phi-\phi_3)-\Delta_2, & \text{for} & \phi>\phi_3,
 \end{array}\right.
\end{equation}
with $\phi_3<4$, the initial conditions must be $\phi(0)\geq4$ even for $D=3$. This implies that there exists no finite instanton solution for $D=4$.

With the previous information about the behavior of instanton solutions in the neighborhood of fixed points, using the so-called qualitative theory of nonlinear dynamical systems \cite{Zeeman1977,Guckenheimer1986} it is possible to construct functions with the same topological and asymptotic properties of the exact solutions of Eq.~\eqref{Eq20}. Thus, it is possible to generalize the results for some classes of equations that are topologically equivalent to those with exact solutions. With these ideas in mind, it is interesting to study Lagrangians where the potential $U(\phi)$ behaves as
\begin{subequations}
\label{Eq26}
\begin{align}
\label{Eq26a}
U(\phi)\simeq\frac{1}{n_1}(\phi-\phi_1)^{n_1}&, \quad\text{near }\phi=\phi_1,\\
\label{Eq26b}
U(\phi)\simeq\frac{1}{n_2}(\phi_2-\phi)^{n_2}+\Delta_2&, \quad\text{near }\phi=\phi_2,\\
 \label{Eq26c}
 U(\phi)\simeq\frac{1}{n_3}(\phi-\phi_3)^{n_3}-\Delta_3&, \quad\text{near }\phi=\phi_3,
\end{align}
\end{subequations}
where $n_1\geq2$, $n_2\geq2$, and $n_3\geq2$. As $n_1$, $n_2$, $n_3$ and the dimension $D$ increase, the probability of the formation of expanding true vacuum bubbles in a space filled with false vacuum approaches zero.

\section{Numerical simulations and discussion}
\label{Numerical}

Consider a system of identical torsional pendula, where consecutive pendula are coupled by a torsion spring and can oscillate with a common axis under the influence of gravity. In the continuum limit, this system can be described by the sine-Gordon equation
\begin{equation}
 \label{Eq27}
 \partial_{tt}\phi-\partial_{xx}\phi+\sin\phi=0,
\end{equation}
where $\phi$ is the deviation angle from the equilibrium configuration. In Eq.~\eqref{Eq27}, time is normalized with respect to the inverse of the natural frequency of oscillations of each pendulum, $1/\omega_o$, and space with respect to the ratio $c_o/\omega_o$, where $c_o$ is the characteristic phase speed of short-wavelength waves in the linear limit \cite{Peyrard2004}. The static kink solution to Eq.~\eqref{Eq27} is the following
\begin{equation}
\label{Eq28}
 \phi_k(x)=4\arctan[\exp(x)],
\end{equation}
which is a stable soliton \cite{Gonzalez1998,Gonzalez2002,Gonzalez2007,GarciaNustes2012,Oliveira1996}. However, notice that this stable configuration is composed of pendula that are away from the stable equilibrium points, namely $\phi=0$ and $\phi=2\pi$. For instance, the pendulum at $x=0$ is at the angle $\phi=\pi$, which for a single pendulum would be a completely unstable position. In the topological soliton solution given by Eq.~\eqref{Eq28}, the pendulum at position $\phi=\pi$ and its neighbors are sustained by the majority of the pendula, which are close to the stable positions $\phi=0$ and $\phi=2\pi$ in the kink configuration.

\begin{figure}[h]
\begin{center}
    \includegraphics[width=3.7in]{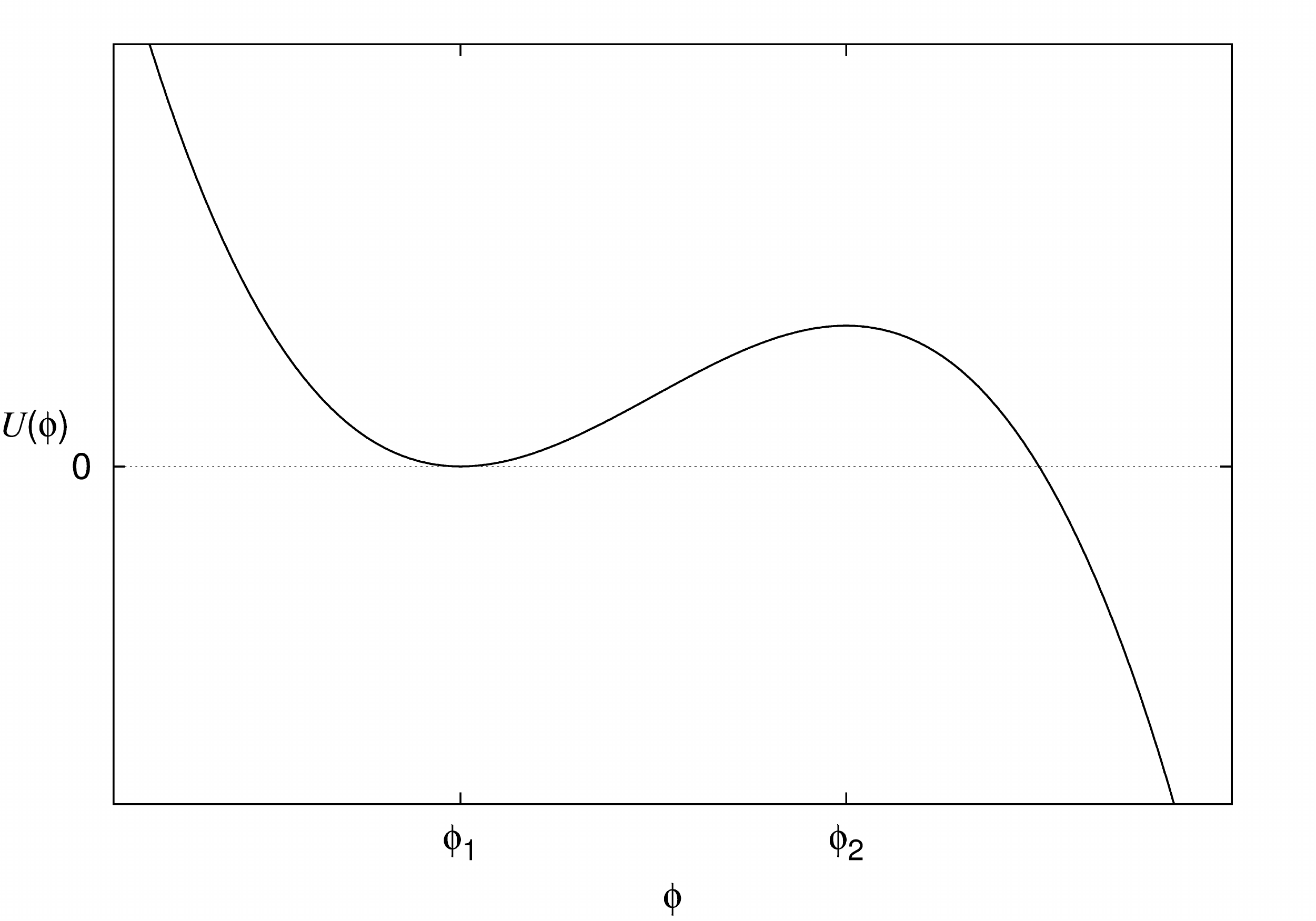}
\end{center}
\caption{Potential $U(\phi)$ of Eq.~\eqref{Eq29} with a minimum at $\phi_1=0$ and a potential barrier at $\phi_2=1$. The initial conditions of Eq.~\eqref{Eq32} represents an initially-at-rest chain of linked balls with some of the balls hanging on the abyss.}
\label{fig:4}
\end{figure}

Now consider a one-dimensional lattice of linked particles which interact with their neighbors via the anharmonic potential
\begin{equation}
\label{Eq29}
U(\phi)=\frac{\phi^2}{2}-\frac{\phi^3}{3},
\end{equation}
which is depicted in Fig.~\ref{fig:4}. This system is a relevant model for the study of fragmentation of long chains \cite{Oliveira1996} and can be described as a chain of balls moving in the potential of Eq.~\eqref{Eq29}. In the continuum limit, the chain dynamics is governed by
\begin{equation}
 \label{Eq30}
 \partial_{tt}\phi-\partial_{xx}\phi=\phi^2-\phi,
\end{equation}
whose exact critical-bubble solution \cite{Oliveira1996} can be obtained as
\begin{equation}
 \label{Eq31}
 \phi(x)=\frac{3}{2}\mbox{sech}^2\left(\frac{x}{2}\right). 
\end{equation}

The potential of Eq.~\eqref{Eq29} has an unstable equilibrium at $\phi_2=1$ between an abyss and a stable equilibrium at $\phi_1=0$. For instance, an initial condition of the form
\begin{equation}
\label{Eq32}
 \phi(x,0)=1.4\,\mbox{sech}^2(x),\quad
 \partial_t\phi(x,0)=0,
\end{equation}
would lead to a dynamics where the chain falls to the potential well at $\phi_1=0$. Equation \eqref{Eq32} corresponds to a configuration where some balls of the chain are situated to the right of the unstable equilibrium point $\phi_2$, as depicted in Fig.~\ref{fig:4}. However, these balls outside the potential well are sustained by those that are inside, and the chain will not fall to the abyss. Larger bubbles like the following
\begin{equation}
 \label{Eq33}
 \phi(x,0)=1.6\mbox{sech}^2\left(\frac{x}{4}\right),
\end{equation}
will grow forever.

\begin{figure*}[tp]
\centerline{\includegraphics[width=6in]{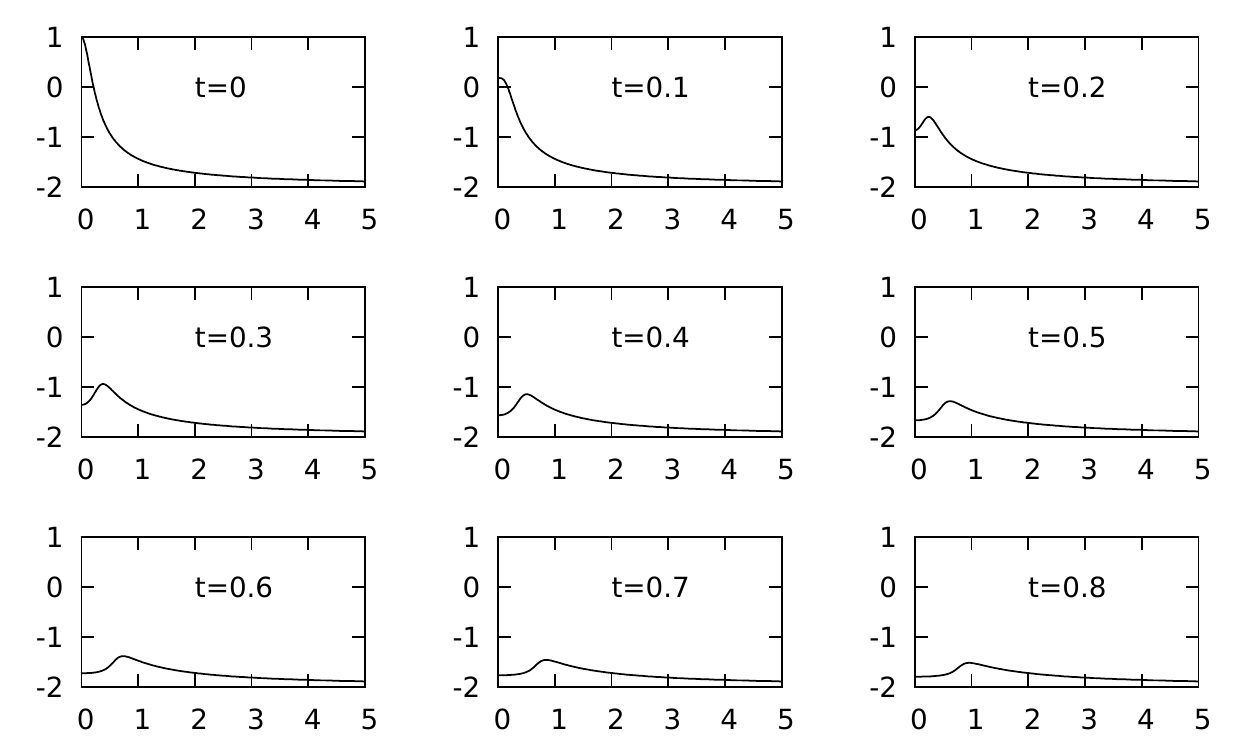}}
\caption{$\phi(r)$ for different times. Collapse of a small bubble of true-vacuum inside a sea of false vacuum.} 
\label{fig:6}
\end{figure*}

\begin{figure*}[tp]
\centerline{\includegraphics[width=6in]{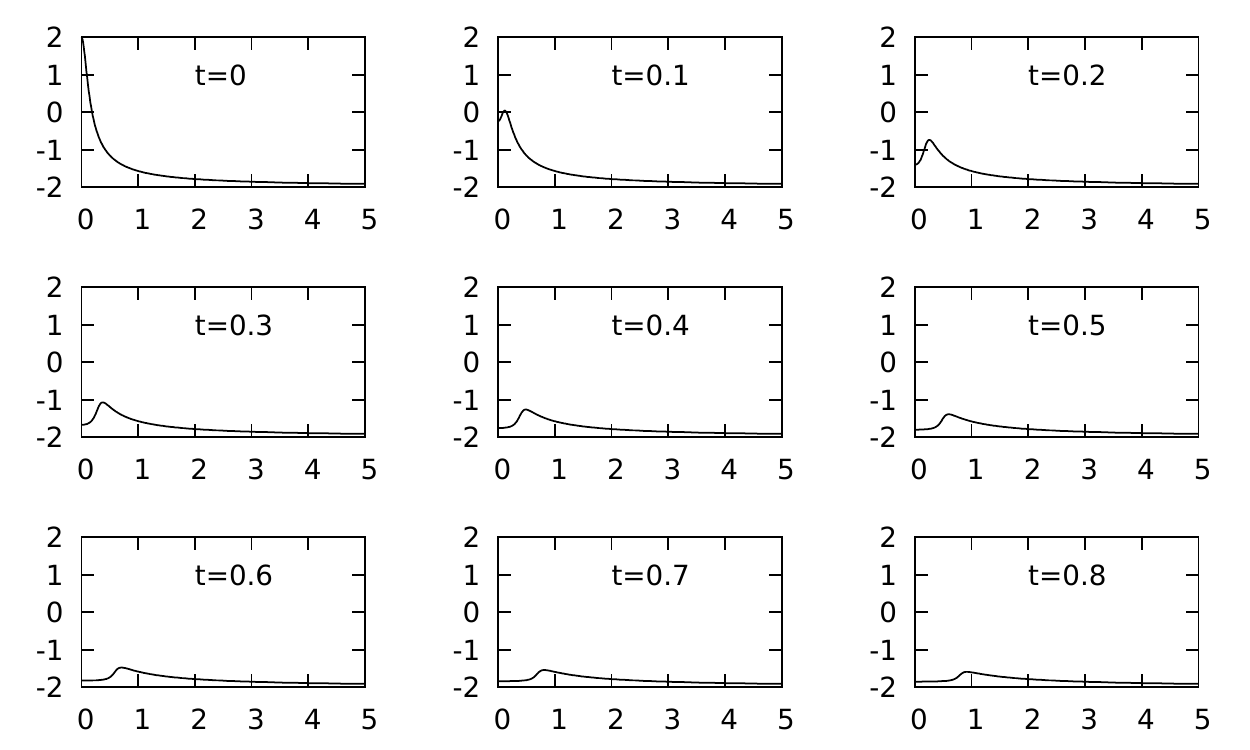}}
\caption{$\phi(r)$ for different times. Collapse of a true-vacuum bubble inside a sea of false vacuum. Notice that $\phi(0)=2$.}
\label{fig:7}
\end{figure*}

\begin{figure*}[tp]
\centerline{\includegraphics[width=6in]{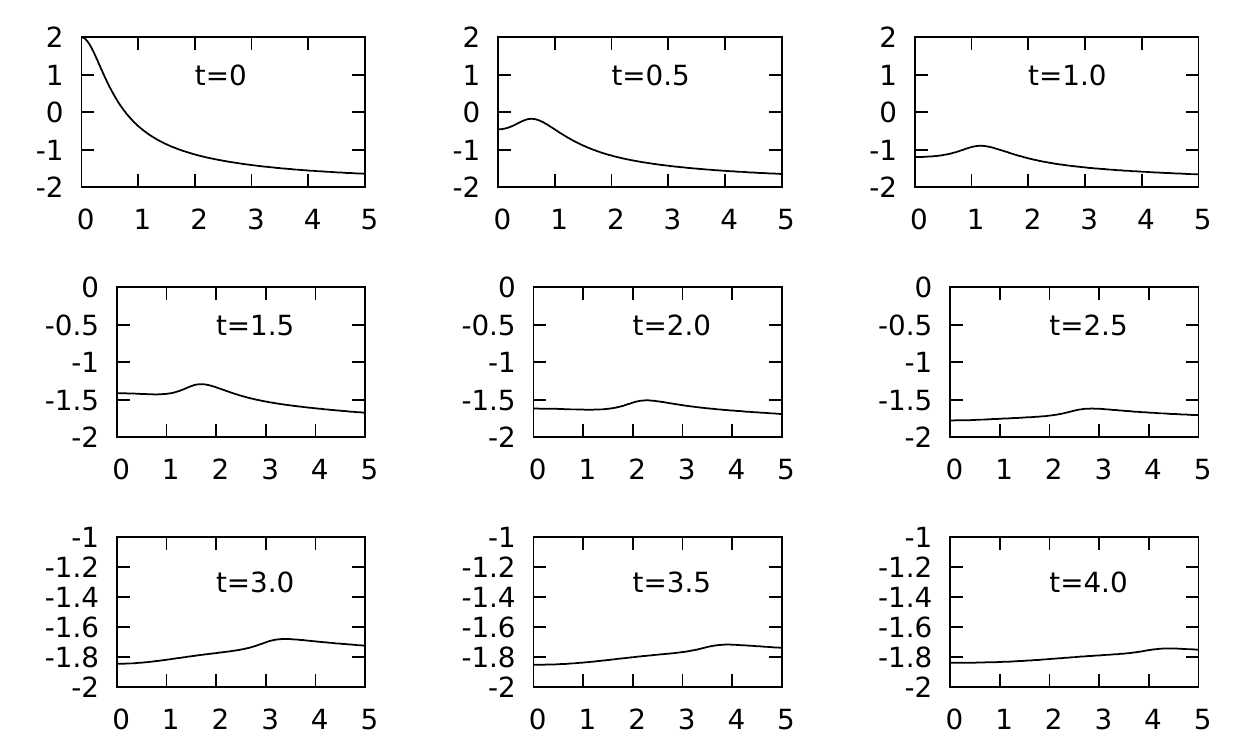}}
\caption{$\phi(r)$ for different times. Collapse of a true-vacuum bubble inside a sea of false vacuum. Notice that the tail of the bubble is thicker.}
\label{fig:8}
\end{figure*}

\begin{figure*}[tp]
\centerline{\includegraphics[width=6in]{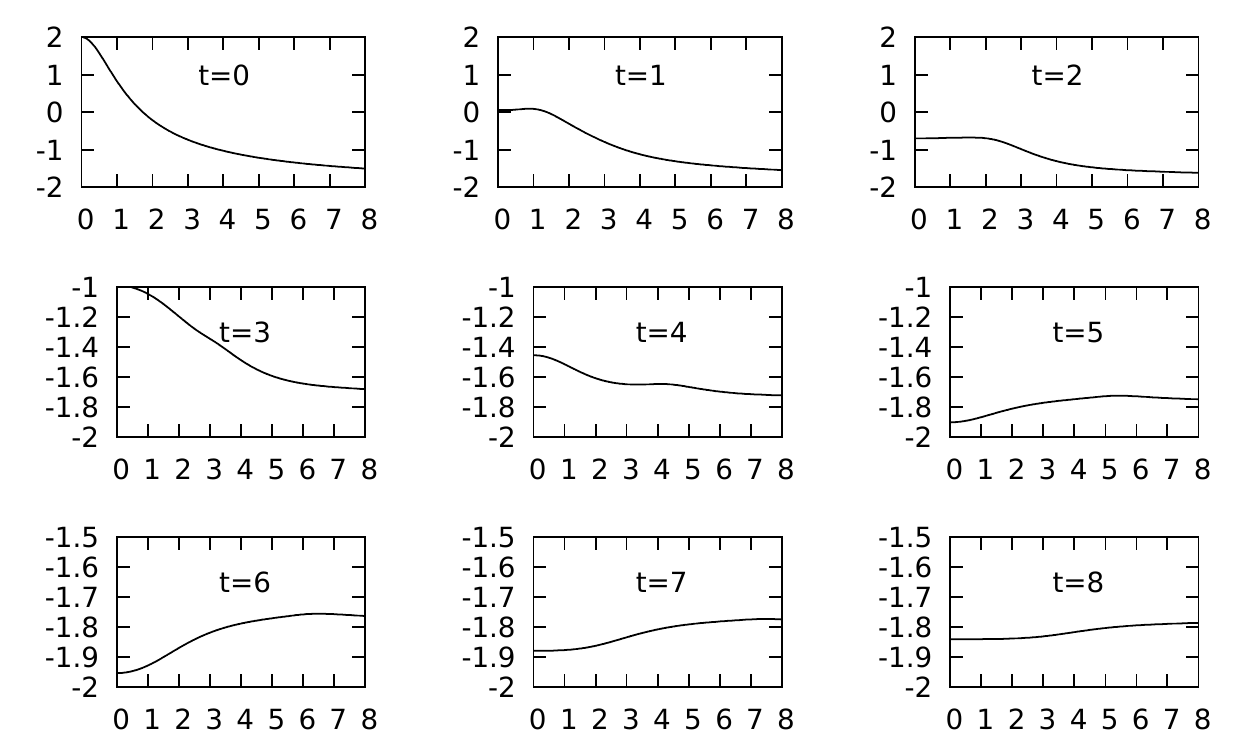}}
\caption{$\phi(r)$ for different times. Collapse of a large true-vacuum bubble inside a sea of false vacuum.}
\label{fig:9}
\end{figure*}

\begin{figure*}[tp]
\centerline{\includegraphics[width=6in]{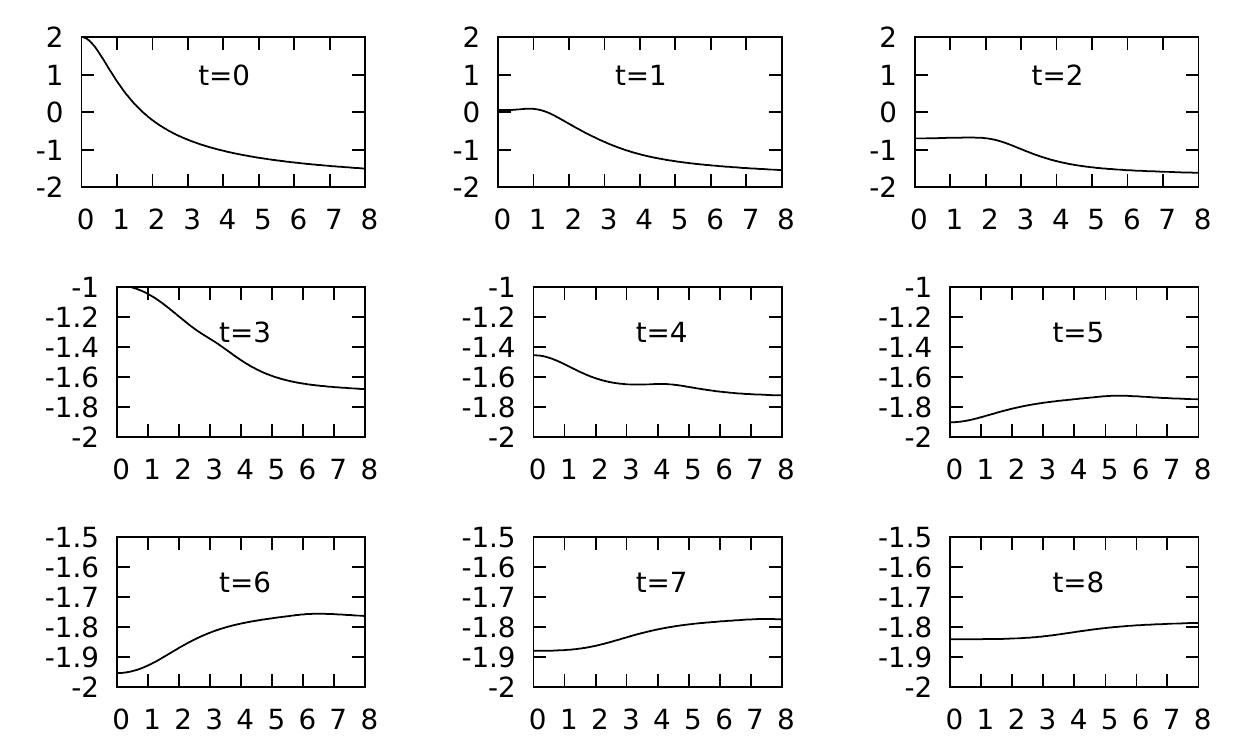}}
\caption{$\phi(r)$ for different times. Collapse of a large true-vacuum bubble inside a sea of false vacuum. Notice that the tail is very thick.}
\label{fig:10}
\end{figure*}

\begin{figure*}[tp]
\centerline{\includegraphics[width=6in]{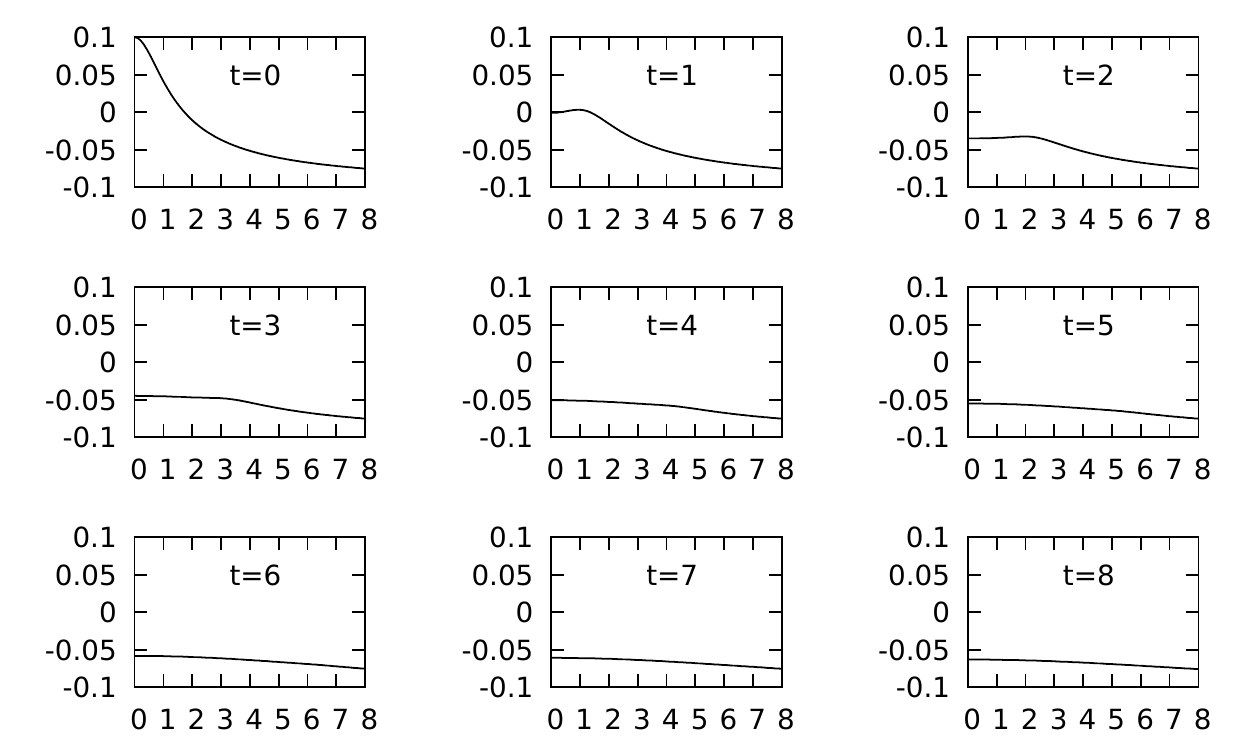}}
\caption{$\phi(r)$ for different times. The initial condition is not an exact solution to the field equation. The values of $\phi$ are very close to the maximum of $U(\phi)$. However, the solution $\protect\phi (\mathbf{r},t)$ does not tend to any of the minima.}
\label{fig:11}
\end{figure*}

\begin{figure*}[tp]
\centerline{\includegraphics[width=6in]{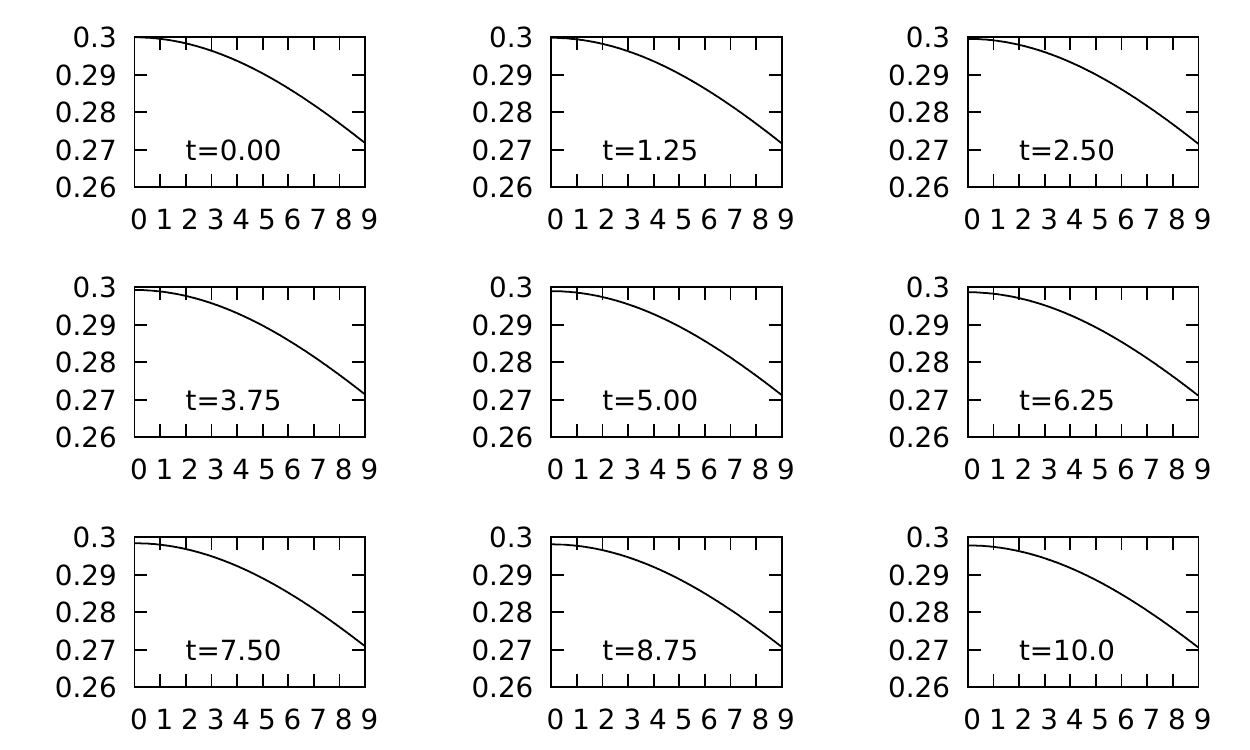}}
\caption{$\phi(r)$ for different times. A stationary bubble in $D=3+1$ dimensions. The action of the bubble is infinite. However, the bubble does not expand.}
\label{fig:12}
\end{figure*}

\begin{figure*}[tp]
\centerline{\includegraphics[width=6in]{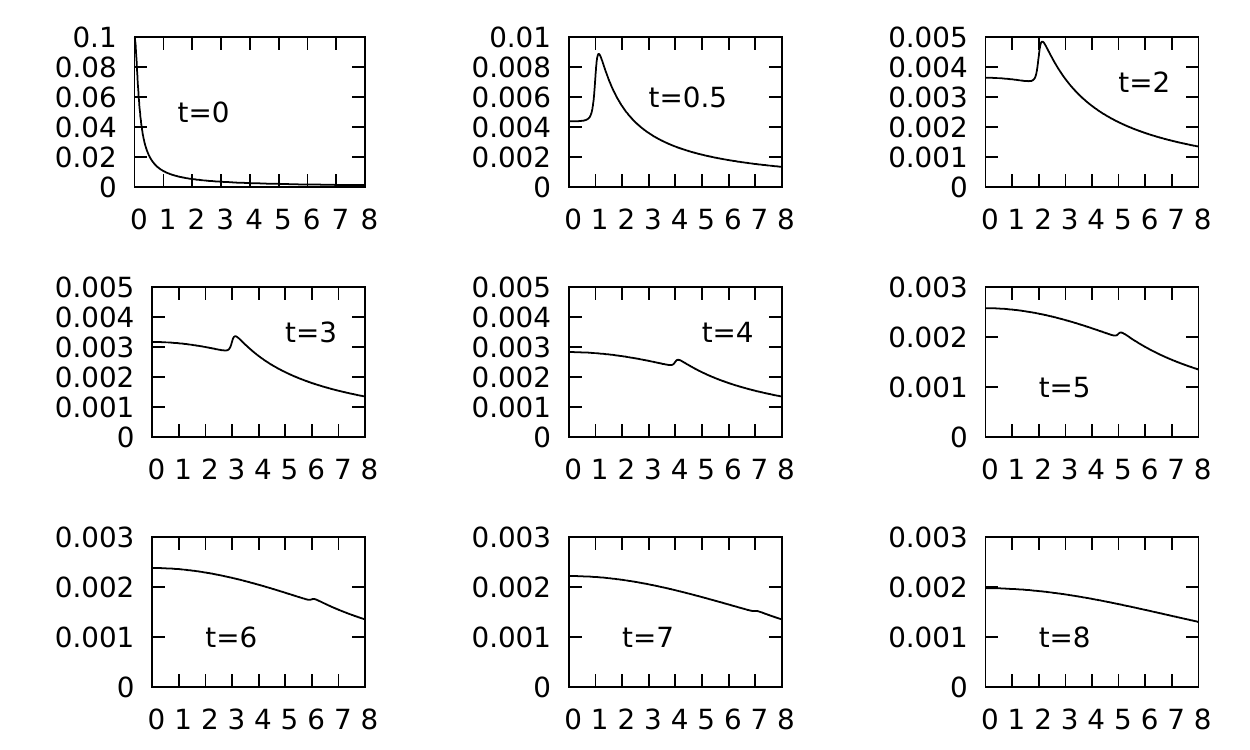}}
\caption{$\phi(r)$ for different times. The initial condition is not an exact solution to the field equation. Initially there are some dynamics and the bubble re-accommodates. Then, the bubble becomes stationary.}
\label{fig:13}
\end{figure*}

\begin{figure*}[tp]
\centerline{\includegraphics[width=6in]{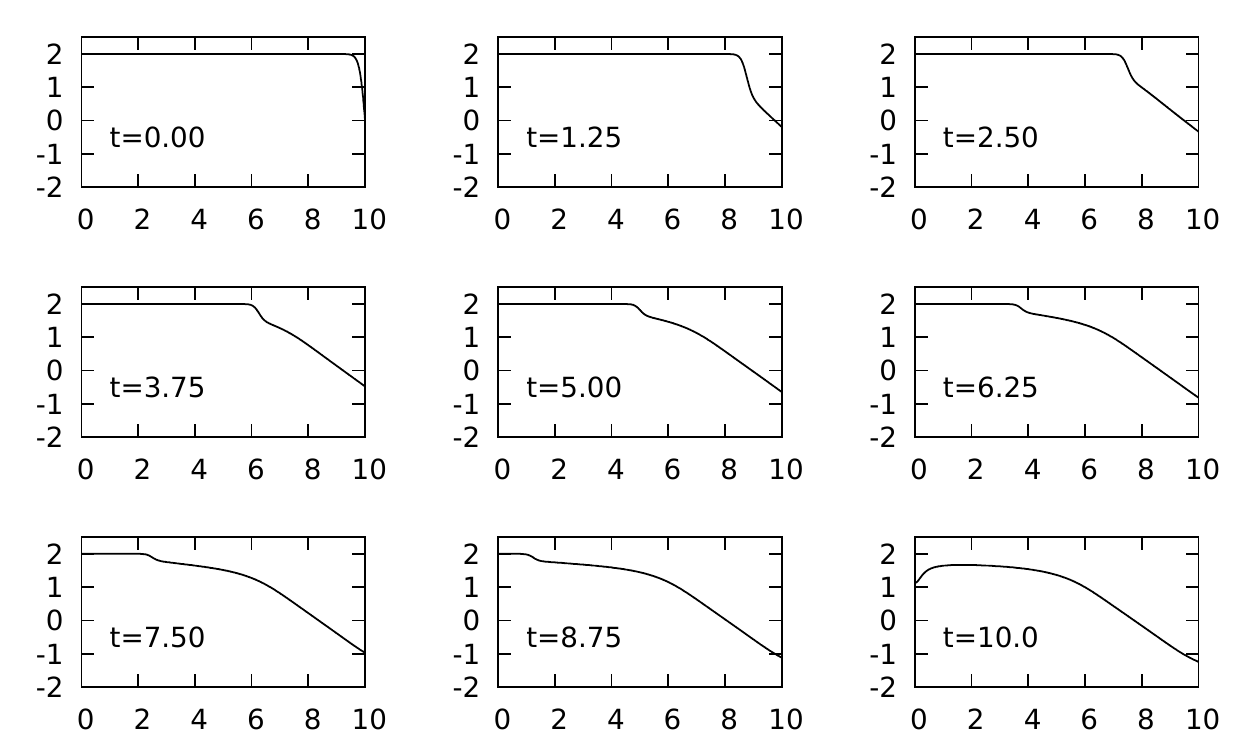}}
\caption{$\phi(r)$ for different times. Dynamics of a very large bubble of true-vacuum inside a Universe of false vacuum. The bubble is almost stationary and long-lived. The bubble does not expand.}
\label{fig:14}
\end{figure*}

Now consider the following equation
\begin{subequations}
\begin{align}
\label{Eq34a}
 &\partial_{tt}\phi-\nabla^2\phi=F(\phi),\\
 &F(\phi)=a\left[\phi(\phi+2)(2+\delta-\phi)\right]^{2n-1},
\end{align}
\end{subequations}
where $n\geq3$. In $D=3+1$ dimensions, the balls that are close to the unstable equilibrium position $\phi_2=0$ are capable to sustain the balls that are arbitrarily away from the equilibrium. Initial conditions like the following
\begin{subequations}
\label{Eq35}
\begin{align}
\label{Eq35a}
 \phi(x,y,z,0)&=\phi_o+\frac{Q}{\sqrt{1+b(x^2+y^2+z^2)}},\\
 \label{Eq35b}
 \partial_t\phi(x, y, z,0)&=0,
\end{align}
\end{subequations}
with $Q>0$ and $b>0$, where all the balls are outside the unstable equilibrium, do not lead to an evolution where the chain falling to the abyss. Figures \ref{fig:6}--\ref{fig:14} show  the results from numerical simulations for
different values of parameters $\phi_0$, $Q$, and $b$. The initial structure is unstable and collapses in time, giving a constant phase $\phi_0$ throughout all space at long times. Indeed, in all such cases, $\phi$ has positive as well as negative values in space for $t=0$. 
The value of parameter $b$ is large enough so that the bubble is strongly localized around $\phi_{max}:=\phi_0+Q$. Thus, as the system evolves in time, the solution falls towards the stable minimum at $\phi=\phi_0$ in all space.

Note that there are bubbles whose action is infinite and are made mostly of true vacuum, but, nevertheless, do not grow. Therefore, they do not expand throughout the Universe converting false vacuum to true. This leads to the conclusion that even finite-action true-vacuum bubbles in a sea of false vacuum will not grow either.

Now consider an initial condition where the majority of the balls are close to the locally stable equilibrium $\phi=\phi_1$. In $D=3+1$ dimensions, the system can sustain any field configuration with the values of $\phi$ arbitrarily far to the right from the unstable equilibrium $\phi=\phi_2$. In $D=3+1$ the link between the balls in the chain is very strong and the balls outside the potential well cannot drag those that are inside. Initial conditions in the form
\begin{subequations}
\label{Eq36}
\begin{align}
\label{Eq36a}
 \notag \phi(x,y,z,0)&=-2+A[\tanh[B(r+d)]\\-&\tanh[B(r-d)]],\\
 \label{Eq36b}
 \partial_t\phi(x, y, z,0)&=0,
\end{align}
\end{subequations}
will not evolve in such a way that the bubble will expand, and the false vacuum will be converted to the true vacuum.

The previous ideas are illustrated in Figures~\ref{fig:6}--\ref{fig:10}, where bubbles made of true vacuum collapse very fast inside a Universe made of false vacuum. However, notice that there is a relaxation process such that the field $\phi$ tends asymptotically to the stable equilibrium $\phi=\phi_1$.

Figures \ref{fig:11} and \ref{fig:13} show a peculiar phenomenon. The initial conditions in such numerical simulations are not exact solutions to the field equation. Initially there is some dynamics and the bubbles re-accommodate. Then the bubbles become stationary. These are high-energy states. These bubbles do not expand.

Figure \ref{fig:12} shows the dynamics of a bubble made completely of  true vacuum. This bubble is stationary and marginally stable. There is a whole continuum zone of bubbles of different amplitudes and different behaviors of the tails. All the bubbles in this zone are stationary 
and unstable solutions to the field equations. A perturbation of the bubble will lead to a different stationary configuration that belongs to the mentioned zone. However, the most remarkable fact is that these bubbles do not expand! Figure \ref{fig:13} shows the re-accommodation of an initial condition that is not an exact solution to the field equations.

Figure \ref{fig:14} shows the dynamics of a very large bubble made of true vacuum inside a Universe of false vacuum. This bubble does not expand neither and can be regarded as a quasi-stationary long-lived structure that could contain new physics and new particles.

Our results suggest that there are no bubbles of true vacuum inside a Universe of false vacuum that will expand. All bubbles of the true vacuum inside a Universe of the false vacuum will collapse.


\section{Discussion on further physical phenomena}
\label{Discussion}

\subsection{Criticality}

Recent theoretical and experimental works \cite{Degrassi2012,Burda2016,EliasMiro2012,Grinstein2016,Sen2015,Burda2015,Ema2016,Hoang2015,Aad2012,Chatrchyan2012,Andreassen2014,Aad2015,Khachatryan2016,Bezrukov2012,Aabound2016} suggest that both $m^2$ and $\lambda$, the two parameters of the Higgs potential, are zero or approximately zero. They happen to be near critical lines that separate the electroweak phase from a different phase of the Standard Model (SM). The multiverse is a fertile and constructive framework that could help us to reformulate in different terms some open questions in fundamental physics. The idea that the parameters of the Higgs potential could be dynamical variables offers a variety of novel approaches to the problems of contemporary physics.

Intriguingly, the Higgs problem can be re-stated in terms of criticality \cite{Giudice1988}. Why are the Higgs parameters so close to the critical point of the electroweak phase transition? The discovery of the Higgs boson at 125 GeV has revealed another potential problem of criticality within the SM \cite{Butazzo2013}: why is the Higgs mass so close to the critical point for a new phase transition? These questions may suggest an underlying dynamics of the Higgs potential 
parameters that drive them towards critical points in the phase diagram.

The Higgs potential could undergo a self-tuning process, with critical boundaries acting as attractors, just as in the mechanism of self-organized criticality \cite{Bak1987}. In the best of scenarios, our Universe is in a minimum of the potential $U(\phi)$. This potential can behave as $U(\phi)\simeq \left|(\phi - \phi_1)^6\right|/6$ when $\phi\simeq \phi_1$. We have shown that if $U(\phi) \simeq \left|(\phi - \phi_1)^{n_1}\right|/{n_1}$, where $n_1 > 2$, then the action of the instanton is infinite. But even if $m^2 \simeq 0$ and $\lambda \simeq 0$, the minimum will be very flat. This leads to the fact that the action of the instanton is arbitrarily large. Self-organized criticality can make the parameters approach the critical values during a dynamical process. Thus, criticality means that the action of the instanton is arbitrarily large. Hence our vacuum is safe.

\subsection{Origin of matter-antimatter asymmetry}

The recent measurement of the Higgs boson mass implies a relatively slow rise of the Standard Model Higgs potential at large scales and a possible second minimum at even larger scales. During inflation, scalar fields, including the Higgs field, may acquire a non-zero vacuum expectation value, which must later relax to the equilibrium value during reheating. In the presence of the time-dependent condensate, the vacuum state can evolve into a state with non-zero particle number. Scientists have shown \cite{Kusenko2015c,Pearce2015} that, in the presence of the lepton number violation in the neutrino sector, the particle production can explain the observed matter-anti\-matter asymmetry of the Universe.

According to Ref.~\cite{Kusenko2015c}, a false vacuum can appear at large values of $\phi$. Our results show that the field can become trapped in the false vacuum. When reheating begins, finite-temperature effects can eliminate this false vacuum and the field rolls down to the present minimum. The stationary structures with large values of $\phi$ and the quasi-stationary long-lived soliton-like bubbles of a different vacuum that are discussed in  this paper can contribute to the acquisition of non-zero vacuum expectation value. Thus, these phenomena can lead to the creation of matter-antimatter asymmetry in the Universe.

\subsection{Confinement}

Let us consider Lagrangians with a potential $U(\phi)$ that possesses two  minima $\phi_1$ and $\phi_3$. Despite the fact that $U(\phi_3) < U(\phi_1)$, in some cases, the state 
$\phi = \phi_1$ is not rendered unstable even by quantum penetration. The Extra-dimensions help to the stabilization of the state $\phi = \phi_1$. This can be considered a mechanism for confinement.

\subsection{Multiverse. String theory landscape of vacua. Extradimensions}

Much of the popularity of the Multiverse was brought about by the discovery of the enormous landscape of possible vacua in string theory \cite{Bousso2000,Freivogel2004}. At present, string theory is the leading candidate for the theory of quantum gravity. As is well-known, the landscape contains about $10^{600}$ different vacua. Accordingly, if our Universe is initially in the false vacuum, it can only tunnel from its false vacuum state to its nearest neighbor that is then the new true vacuum state. The latter would be arranged according to the anthropic principle. Consequently, the extraordinarily fine-tuning of this potential would be due to the anthropic selection.

We conclude that the dynamics in the landscape is described by localization using simpler arguments than those presented in Ref.~\cite{Mersini}. The fine-tuning in our scenario is produced by the properties of our Universe discussed above. Once we are in the vacuum of our Universe, the mechanism that we have described above inhibits the escape and explains why the solutions are localized around a vacuum. There are no instanton solutions in our Universe.

The space-time dimension $D$ plays an important role in the mechanism that makes our vacuum stable. Thus, the existence of extra-dimensions is not in contradiction with the great stability of our Universe. If the soliton-like solutions that we have found are considered classical string solutions, then the formulation of string theory is straightforward.

\subsection{Dark energy and dark matter}

The stationary solutions shown in Figs.~\ref{fig:11}--\ref{fig:13} close to the maximum of $U(\phi)$ (see Fig.~\ref{fig:2}) possess enormous energy. They can be good candidates for dark energy.  Inside the true-vacuum bubbles represented in Fig.~\ref{fig:14}, there are new particles and new physics.

In 2010, Meng Su and coworkers \cite{Su2010} discovered enormous galactic bubbles that expose dark matter \cite{Su2011,Slatyer2014}. These bubbles are strange structures very similar to the bubbles discussed in Fig.~\ref{fig:14}. These structures are sometimes called Fermi bubbles \cite{Su2011}, although there are no theories that explain them and it seems that dark  matter can be related to such bubbles. Indeed,  looking at the Fermi bubbles near the galactic center, M. Su and coworkers have found a promising signal that could be associated with dark matter. This is a very concrete case where studies of the Fermi bubbles uncovered something that may be related to dark matter.

The existence of long-lived bubbles of true vacuum containing new physics, such as the bubble shown in Fig.~\ref{fig:14}, may be related to galactic bubbles exposing dark matter. Thus, the findings of M. Su and coworkers may be an indirect experimental confirmation of the results discussed in this article.

\section{Conclusions}
\label{Conclusions}

We have investigated the soliton-like bubbles and the instanton solutions in vacuum dynamics. Recent experimental evidence indicates that the two parameters of the Higgs potential are zero or approximately zero. We have shown that this near criticality leads to an arbitrarily large action of the instanton solution, suggesting that the probability of our vacuum decay is exponentially small. We have provided new evidence that supports a process for the origin of the matter-antimatter asymmetry in our Universe proposed recently by other scientists \cite{Kusenko2015c,Pearce2015}. The Extra-dimensions can play an important role in the stabilization of the ``false'' vacuum. We review a possible straightforward formulation of string theory, the multiverse, the string theory landscape, and the extra-dimensions using our results. We have presented a new mechanism for confinement and we utilize our solutions to introduce some hypotheses about dark energy and dark matter. We examine the discovery of giant galactic bubbles exhibiting dark matter, made by a team of scientists working at the Harvard-Smithsonian Center for astrophysics \cite{Su2010,Su2011,Slatyer2014}. The discovery of such bubbles may be an indirect experimental confirmation of our results.

\begin{acknowledgements}
J.F.M. acknowledges partial financial support of Universidad de Santiago de Chile through POSTDOC\_DICYT project number 042031GZ\_POSTDOC and ANID FONDECYT/POSTDOCTORADO/3200499.
\end{acknowledgements}

%
 \section*{Conflict of interest}

The authors declare that they have no conflict of interest.



\end{document}